\documentclass[twocolumn,showpacs,aps,prl,superscriptaddress,floatfix,nofootinbib]{revtex4-2}
\usepackage{graphicx}
\usepackage{verbatim}
\usepackage{dcolumn}
\usepackage{amsmath}
\usepackage{epsfig}
\usepackage{subfigure}
\usepackage{soul}
\usepackage{xcolor}
\usepackage[switch]{lineno}
\usepackage{multirow}
\usepackage[colorlinks=true, linkcolor=blue, urlcolor=black]{hyperref}

\newcommand{\BaBarYear}      {23}
\newcommand{\BaBarNumber}    {01}
\newcommand{\BaBarType}      {PUB}
\newcommand{\SLACPubNumber}  {17720}

\input{babar_sym.tex}

\def\MeVcc{\mevcc}
\def\GeVcc{\gevcc}

\def\figurebox#1#2#3{%
    \def\arg{#3}%
    \ifx\arg\empty
    {\hfill\vbox{\hsize#2\hrule\hbox to #2{\vrule\hfill\vbox to #1{\hsize#2\vfill}\vrule}\hrule}\hfill}%
    \else
    {\hfill\epsfbox{#3}\hfill}%
    \fi}

\begin{document}

\pagestyle{plain}

\begin{flushleft}
\babar-\BaBarType-\BaBarYear/\BaBarNumber \\
SLAC-PUB-\SLACPubNumber\\
%arXiv:XXXX.XXXX [hep-ex]\\

%BAD 3041, version 3
\end{flushleft}

%\title{{\large \bf Search for Baryogenesis and Dark Matter in \B Meson Decays at \babar}}
\title{{\large \bf Search for \texorpdfstring{$B$}{B} Mesogenesis at \texorpdfstring{\babar}{BABAR}}}

\author{J.~P.~Lees}
\author{V.~Poireau}
\author{V.~Tisserand}
%\affiliation{Laboratoire d'Annecy-le-Vieux de Physique des Particules (LAPP), Universit\'e de Savoie, CNRS/IN2P3,  F-74941 Annecy-Le-Vieux, France}
\author{E.~Grauges}
%\affiliation{Universitat de Barcelona, Facultat de Fisica, Departament ECM, E-08028 Barcelona, Spain }
\author{A.~Palano}
%\affiliation{INFN Sezione di Bari, I-70126 Bari, Italy}
\author{G.~Eigen}
%\affiliation{University of Bergen, Institute of Physics, N-5007 Bergen, Norway }
\author{D.~N.~Brown}
\author{Yu.~G.~Kolomensky}
%\affiliation{Lawrence Berkeley National Laboratory and University of California, Berkeley, California 94720, USA }
\author{M.~Fritsch}
\author{H.~Koch}
%\affiliation{Ruhr Universit\"at Bochum, Institut f\"ur Experimentalphysik 1, D-44780 Bochum, Germany }
% ...
%\affiliation{University of Wisconsin, Madison, Wisconsin 53706, USA }
\collaboration{The \babar\ Collaboration}
\noaffiliation

\begin{abstract}
A new mechanism has been proposed to simultaneously explain the presence of dark matter and the matter-antimatter asymmetry in
the universe. This scenario predicts exotic $B$ meson decays into a baryon and a dark sector anti-baryon ($\psi_D$) with branching fractions
accessible at $B$ factories. We present a search for $B \rightarrow \Lambda \psi_D$ decays using data collected by the \babar\ experiment
at SLAC. This reaction is identified by fully reconstructing the accompanying $B$ meson and requiring the presence of a single $\Lambda$
baryon in the remaining particles. No significant signal is observed, and bounds on the $B \rightarrow \Lambda \psi_D$ branching fraction are derived in the range $0.13 - 5.2\times 10^{-5}$ for $1.0 < m_{\psi_D} < 4.2 \GeVcc$. These results set strong constraints on the parameter space allowed by the theory.
\end{abstract}


\maketitle

\setcounter{footnote}{0}

The nature of dark matter (DM) and the baryon asymmetry of the Universe (BAU) are perhaps two of the deepest mysteries of modern particle physics. The latest cosmological observations reveal that visible matter only accounts for about 15\% of the matter in the universe, while the remaining 85\% is constituted by DM~\cite{planck01, planck02}. Beyond its coupling to gravity, the particle properties of DM remain to be elucidated, and many models have been proposed to explain the observed abundance. The visible matter density is equally mysterious, as cosmology predicts a universe born with equal amount of matter and antimatter. A dynamical mechanism, baryogenesis, is required to produce an initial excess of baryons over anti-baryons consistent with cosmic microwave background and big-bang-nucleosynthesis measurements~\cite{pdg2018, bbnReview}.

\begin{figure}[htb]
 \includegraphics[width=0.45\textwidth]{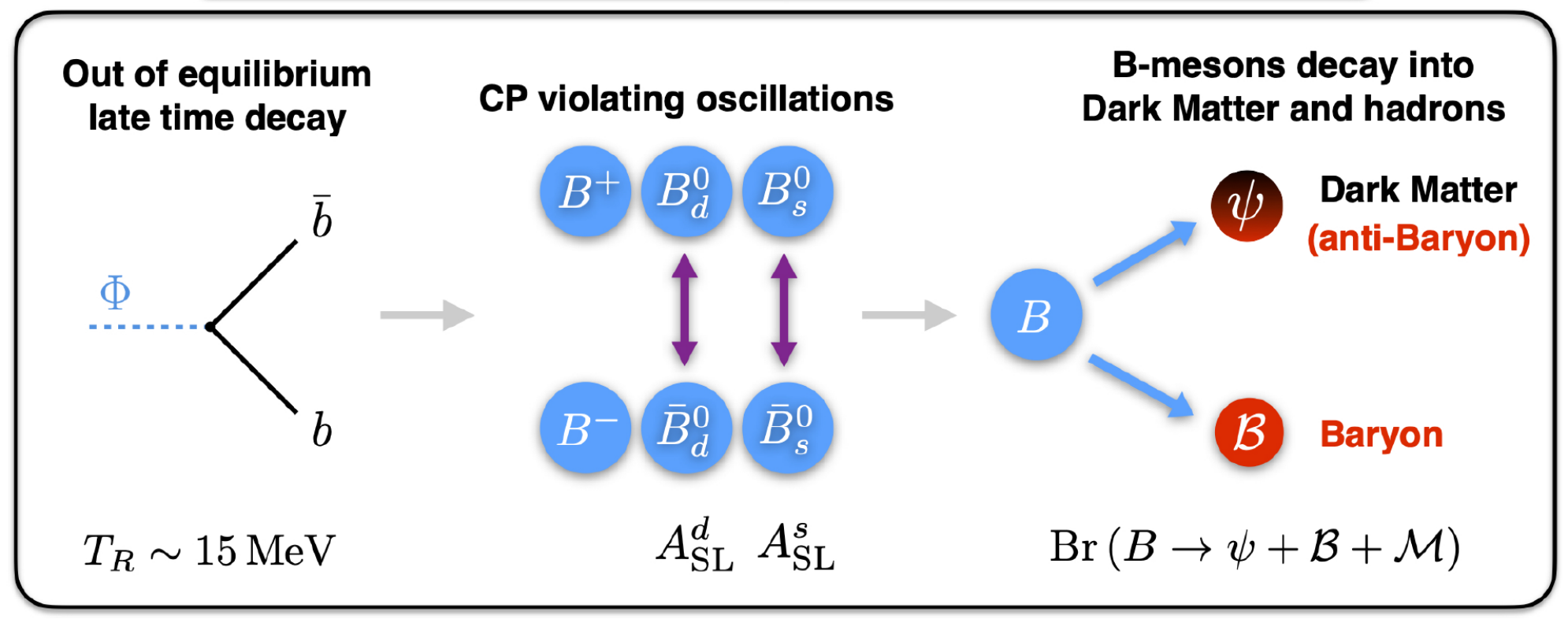}
 \caption{Illustration of the $B$ Mesogenesis mechanism. Figure taken from Ref.~\cite{Alonso-Alvarez:2021qfd}.}
 \label{fig:bDMmechanism}
\end{figure}

The $B$-Mesogenesis scenario has been recently proposed~\cite{nelson, Alonso-Alvarez:2021qfd} to simultaneously explain the DM abundance and
the BAU. This model introduces several new fields, including a light dark-sector anti-baryon and a new TeV-scale
color-triplet bosonic mediator. The baryogenesis mechanism relies on the out-of-thermal-equilibrium production of $b$ and $\bar b$ quarks in the early universe through the decay of a massive, long-lived scalar field $\Phi$, as illustrated in Fig.~\ref{fig:bDMmechanism}. A fraction of these quarks
hadronizes into $B^0$ and $\bar B^0$ mesons, which undergo \CP-violating oscillations before decaying into a baryon $\cal{B}$, a dark-sector anti-baryon $\psi_D$, and any number of additional light mesons $\cal{M}$. As a result, matter-antimatter asymmetries are generated in the visible and dark sectors with equal but opposite magnitudes, keeping the total baryon number conserved.

We note that our search is strongly related in its experimental signature to a recently proposed~\cite{abi} search for supersymmetry in $B$ meson decay to a baryon and undetected light neutralino.

The baryon asymmetry is determined by the charge asymmetry in semi-leptonic $B$ decays, which specifies the level of \CP violation in mixing
in the $B^0 - \bar{B}^0$ system, and the branching fraction of the inclusive $B \rightarrow \psi_D \cal{BM}$ decays. The $B$-Mesogenesis mechanism would imply a robust lower bound on the total branching fraction $\text{BR}(B \rightarrow \psi_D {\cal BM}) > 10^{-4}$~\cite{Alonso-Alvarez:2021qfd}. Constraints on exclusive $B^0 \rightarrow \psi_D {\cal B}$ decays are calculated using phase-space considerations for different baryons~\cite{Alonso-Alvarez:2021qfd}. The results depend on the effective operators ${\cal O}_{i,j} = \psi_D b i j$ mediating the decay, where $i=u,c$ and $j=d,s$ specify the quark content. The ratio of exclusive to inclusive branching fractions ranges from about 1 to $100\%$, depending on the $\psi_D$ mass. Furthermore, bounds on inclusive $b$ decays with missing energy~\cite{ALEPH:2000vvi}, searches for TeV-scale color-triplet scalars at the LHC~\cite{CMS:2019zmd,ATLAS:2020syg}, and dark matter stability require $0.94 < m_{\psi_D} < 3.5 \GeVcc$~\cite{Alonso-Alvarez:2021qfd}. At present, the best constraints on this scenario arise from a measurement of the exclusive $B \rightarrow \psi_D \Lambda$ decay by the Belle Collaboration~\cite{belle} excluding branching fractions larger than $\sim (2-3) \times 10^{-5}$ for $m_{\psi_D} > 1.0 \GeVcc$.

We report herein a search for the decay $B^0 \rightarrow \psi_D \Lambda$ in the mass range
$1.0 < m_{\psi_{D}} < 4.2 \GeVcc$. The analysis is based on 398.5 $\rm fb^{-1}$ of data collected at the $\Y4S$ resonance with
the \babar\ detector at the PEP-II $e^+e^-$ storage ring at SLAC, corresponding to $4.35 \times 10^8$ $B\bar{B}$ pairs~\cite{pepii}.
The \babar\ detector is described in detail elsewhere~\cite{babar1,babar2}. %To avoid any experimental bias, a small sample corresponding to 8\% of the data
An additional 32.5~$\rm fb^{-1}$ of data are used to optimize the analysis strategy and are subsequently discarded. The remaining data
are not examined until the analysis procedure is finalized.

Simulated signal events are created using the EVTGEN~\cite{evtgen} Monte Carlo (MC) event generator. Eight different samples, each with a different $\psi_D$ mass, are generated. The mass values range from 1 to $4.2~\GeVcc$. The background is studied with samples of inclusive $e^+e^- \rightarrow B\bar{B}$ decays (EVTGEN) and continuum $e^+e^- \rightarrow q\bar{q}$ events
with $q=u,d,s,c$ (JETSET~\cite{jetset}). The detector response is simulated with Geant4~\cite{geant4_1, geant4_2}.

Since dark-sector particles escape undetected, we identify the signal by fully reconstructing the second $B$ meson ($B_{\rm tag}$) from hadronic decay
modes, and require the presence of a single $\Lambda$ baryon among the remaining particles. The $\psi_D$ is identified as the system
recoiling against the $B_{\rm tag}$ and $\Lambda$ candidates. Hadronic $B$ meson decays proceed mostly through charmed mesons, and the $B_{\rm tag}$
candidate is reconstructed via the decays $B \to S X$ by a hierarchical algorithm that combines a ``seed'' meson $S$, such as $D^{(*)0}$, $D^{(*)\pm}$, $D_{s}^{*\pm}$, or $J/\psi$, with a hadronic system $X$ containing up to five kaons and/or pions with total charge 0 or $\pm1$~\cite{hrm01}. The selection of $B_{\rm tag}$ candidates
is based on two kinematic variables: the energy difference $\Delta E = E_{\rm beam} - E_{\rm tag}$ and the beam-energy-substituted mass $m_{\rm ES} = \sqrt{E_{\rm beam}^2 - p_{B_{\rm tag}}^2}$, where $E_{\rm tag}$ and $p_{B_{\rm tag}}$ are the energy and momentum of the $B_{\rm tag}$ candidate in
the $e^+e^-$ center-of-mass frame, and $E_{\rm beam}$ is the beam energy in the same frame.
%At this stage of the analysis, $B_{\rm tag}$ candidates with $-0.20 \GeV <\Delta E < 0.20 \GeVcc$ and $5.20 \GeV < m_{\rm ES} < 5.30 \GeVcc$ are selected

\begin{figure}[htb]
  \includegraphics[width = 0.45\textwidth]{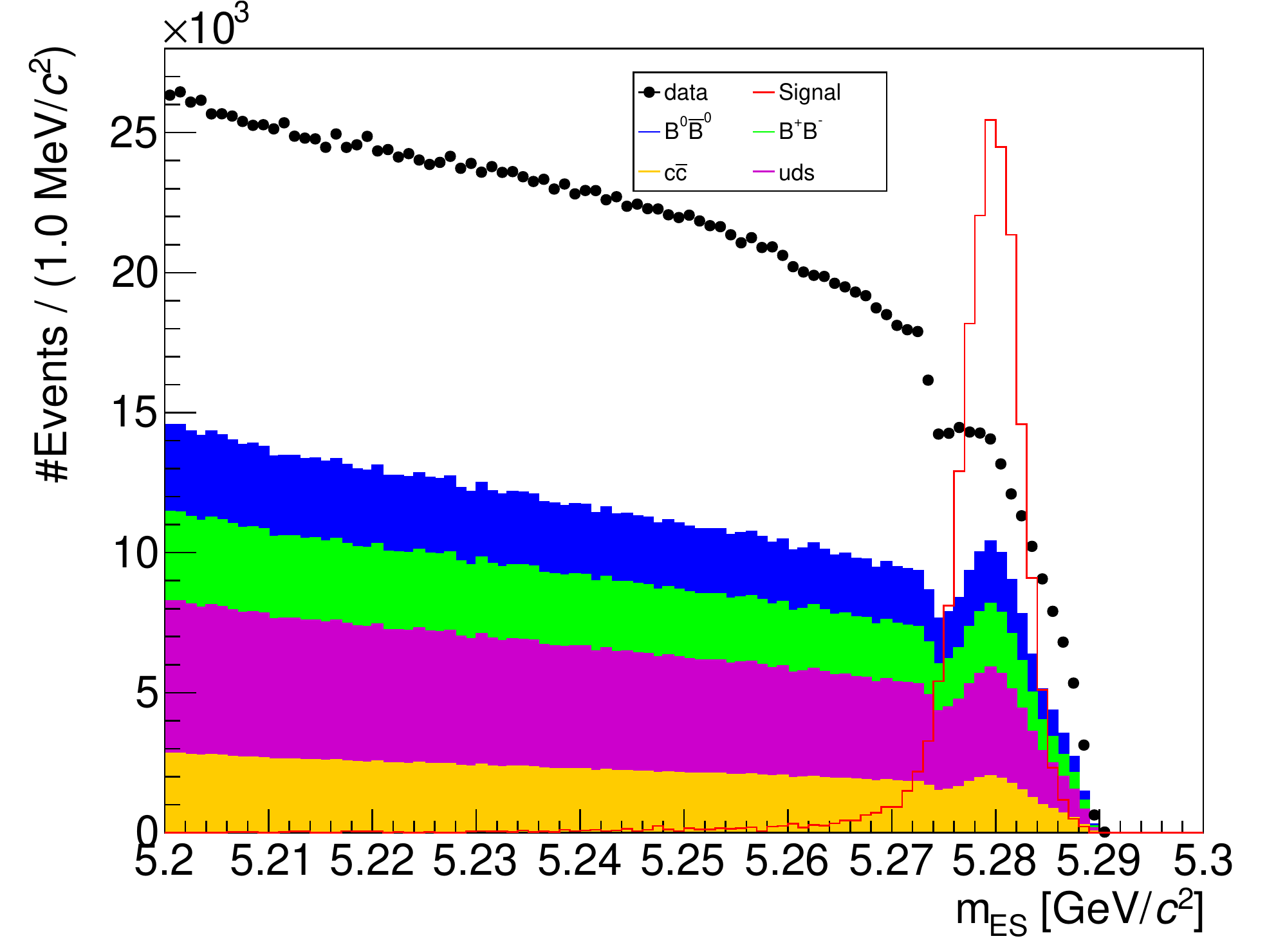}
  \includegraphics[width = 0.45\textwidth]{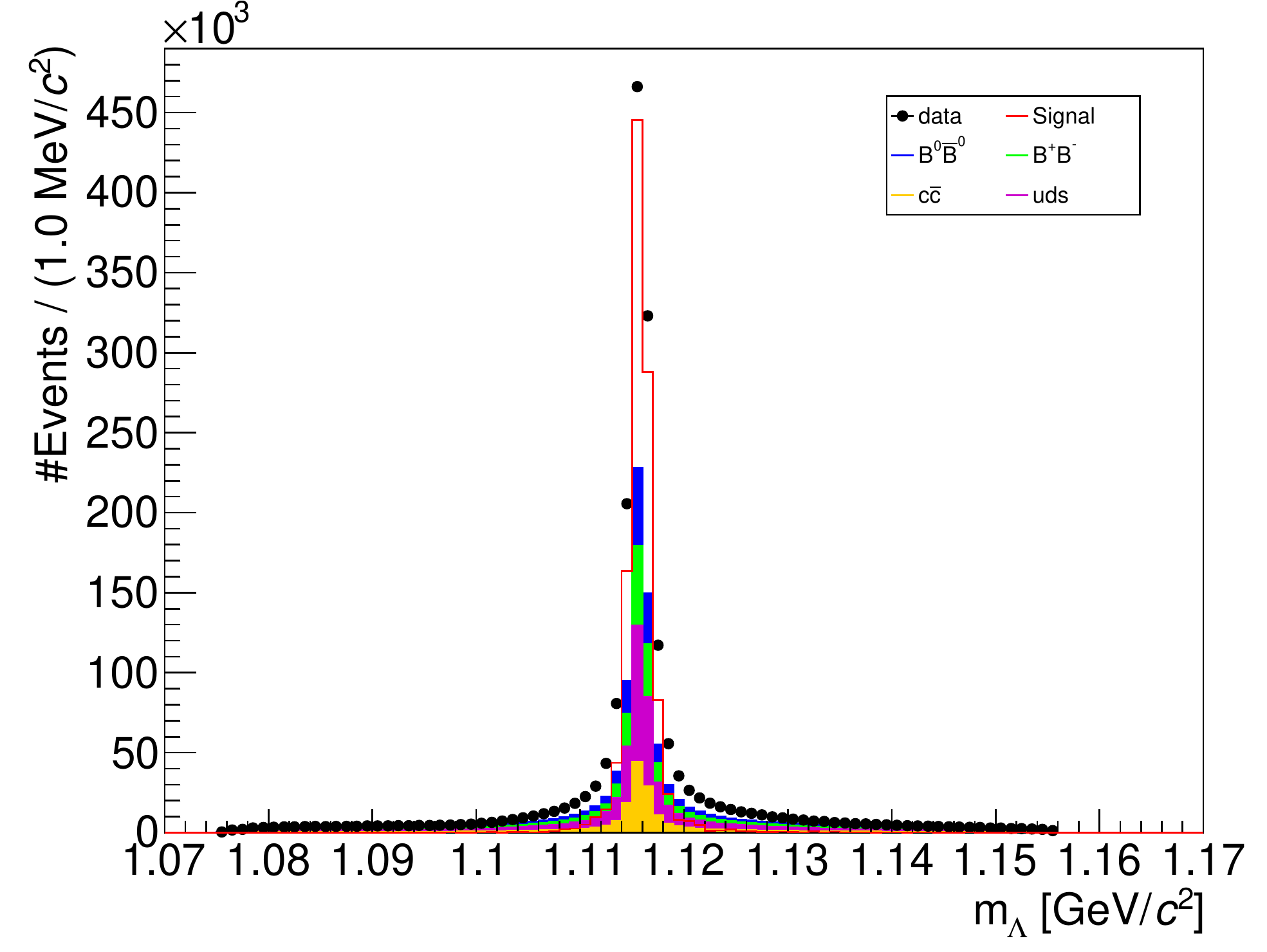}
  \caption{Distribution of (top) the energy-substituted mass, $m_{\rm ES}$, and (bottom) the reconstructed $\Lambda$ mass, $m_\Lambda$,
           for data (points), signal MC for $m_{\psi_{D}} = 2.0 \GeVcc$ (red histogram) and inclusive background MC predictions
	   (stacked histograms). The normalization of the signal events is arbitrary.}
  \label{fig:preDist}
\end{figure}

The remaining particles are associated with the signal $B \rightarrow \Lambda \psi_D$ candidate ($B_{\rm sig}$). The $\Lambda$ candidates are
reconstructed as pair of oppositely charged tracks identified as a proton and a pion. A kinematic
fit is performed on the $\Lambda$ candidate, constraining the two tracks to originate from the same point in space and requiring the momentum vector to point back to the beam interaction region. The $\Lambda$ flight length is calculated as the distance between the primary interaction
point and the secondary decay vertex. The flight length significance, defined as the flight length divided by its uncertainty, must be greater
than 1.0. If more than one combination of $B_{\rm tag}$ and $\Lambda$ candidates is found, the one with the smallest $\chi^2$ is selected. After
reconstructing the $B_{\rm tag}$ and $\Lambda$ candidates, no additional track must be present in the event. The distributions of $m_{\rm ES}$ and the reconstructed $\Lambda$ mass ($m_\Lambda$) are shown in Fig.~\ref{fig:preDist}. We require events to satisfy
$5.27 <m_{\rm ES} < 5.29 \GeVcc$ and $1.110 < m_\Lambda< 1.121 \GeVcc$.

\begin{figure}[ht]
\includegraphics[width = 0.47\textwidth]{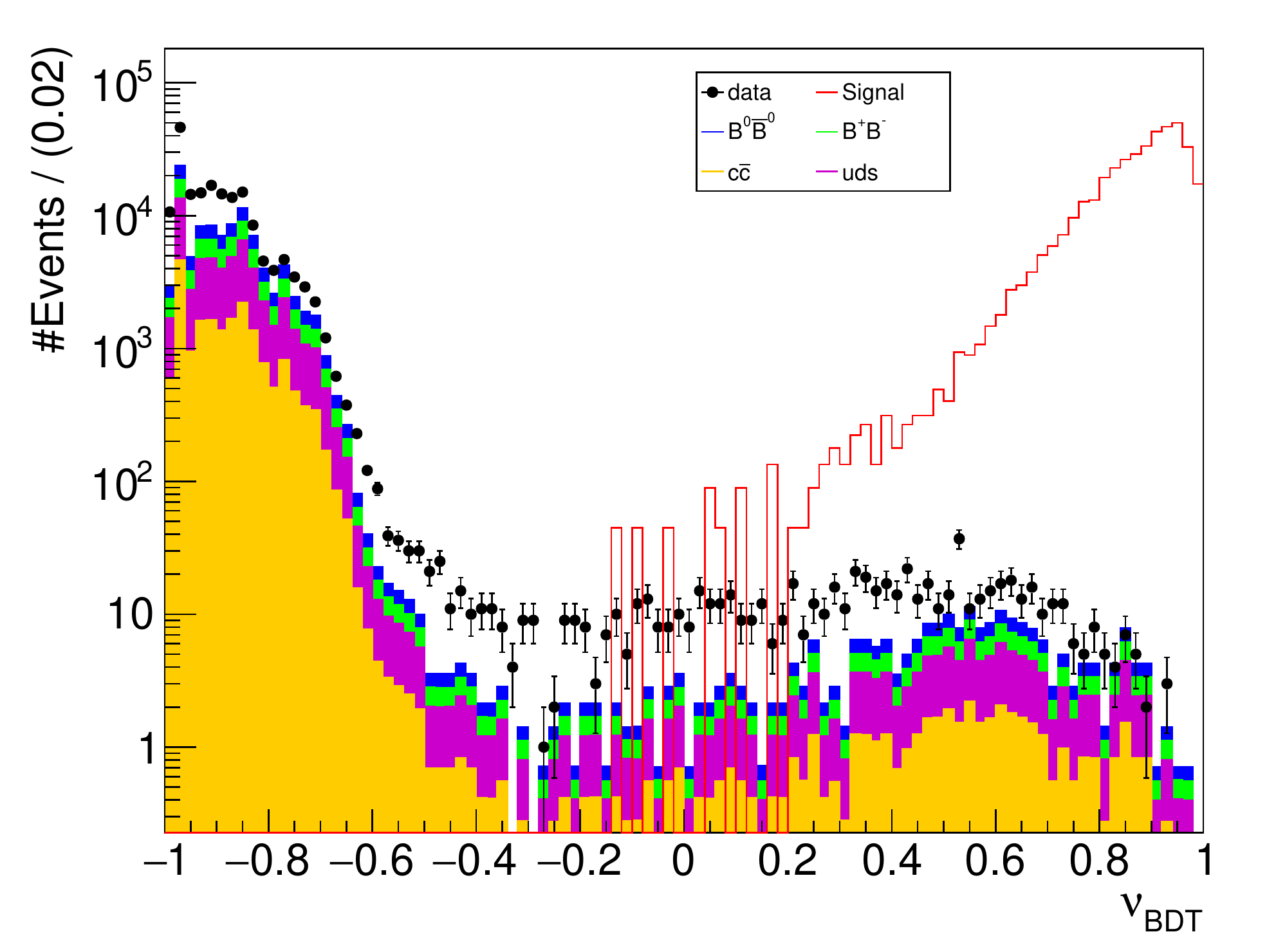}
 \caption{The distribution of the BDT score after applying all other selection criteria for data (points), signal MC for $m_{\psi_{D}} = 2.0 \GeVcc$
 (red histogram) and inclusive background MC predictions (stacked histograms). The normalization of the signal events is arbitrary.}
 \label{fig:BDTscore}
\end{figure}

\begin{figure}[ht]
 \includegraphics[width = 0.47\textwidth]{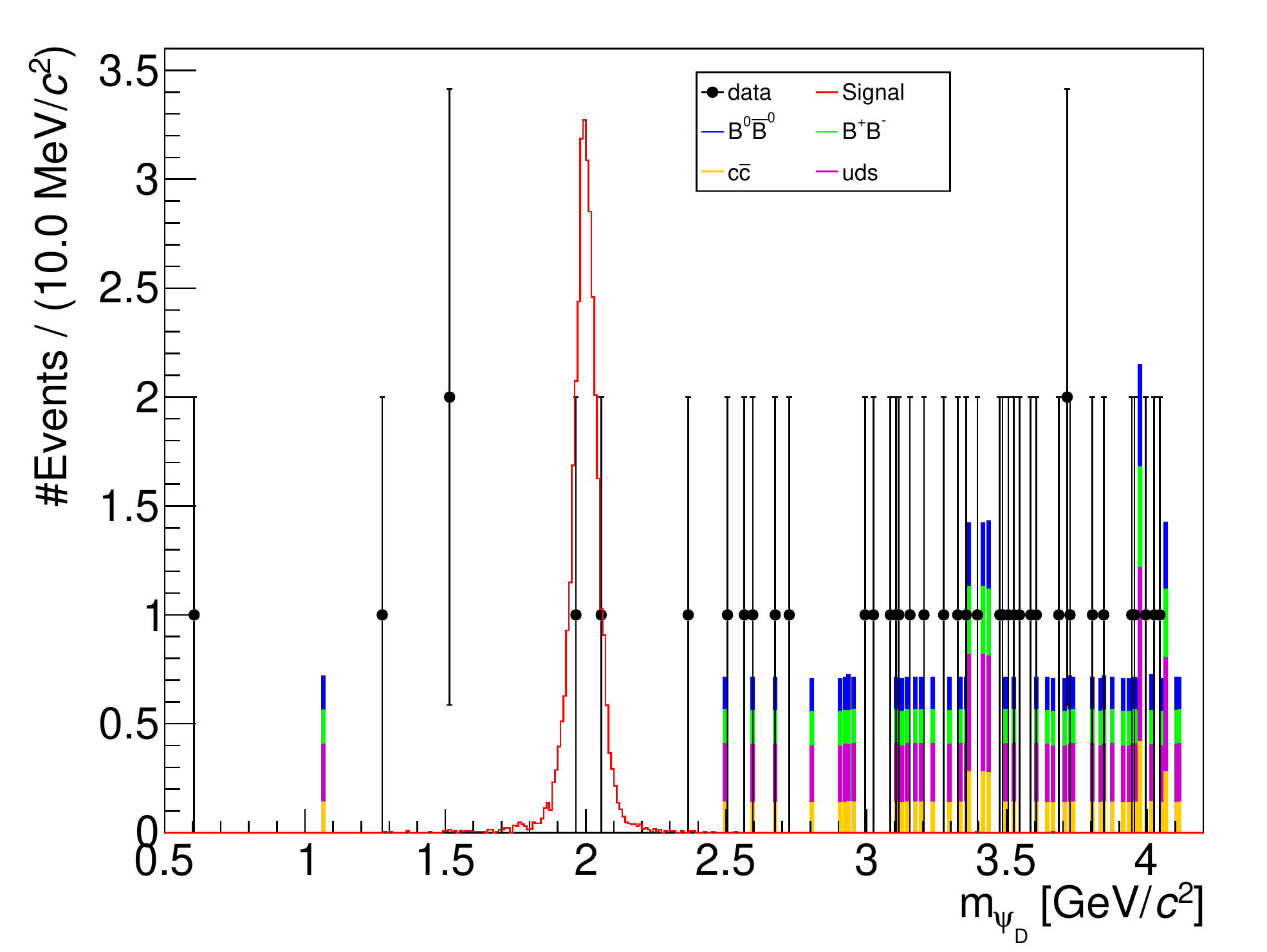}
 \caption{The distribution of the $\psi_D$ mass ($m_{\psi_D}$) after applying all selection criteria for data (points), signal MC for
 $m_{\psi_{D}} = 2.0 \GeVcc$ (red histogram) and inclusive background MC predictions (stacked histograms). The normalization of the signal events is arbitrary.}
 \label{fig:fnlEvts}
\end{figure}

A multivariate selection using boosted decision trees (BDT)~\cite{bdt} is used to further increase the signal purity. The BDT includes the following
variables: $m_{\rm ES}$ and $\Delta E$ of the $B_{\rm tag}$ candidate; information about the $B_{\rm tag}$ hadronic decay
channel and its purity, defined as the fraction of correctly reconstructed $B_{\rm tag}$ candidates for a given decay mode \cite{bfactory}; the magnitude of the $B_{\rm tag}$ thrust vector, defined as the sum of the magnitudes of the momenta of all tracks and calorimeter clusters projected onto the thrust axis \cite{bfactory};
the $B_{\rm sig}$ momentum vector in the laboratory frame, inferred from the initial beam electrons and the recoiling $B_{\rm tag}$; the number of calorimeter clusters associated with $B_{\rm sig}$; the total neutral energy associated with $B_{\rm sig}$; the number of $\pi^{0}$ candidates associated with $B_{\rm sig}$; the $\Lambda$ flight length
significance; the $\chi^2$ of the kinematic fit performed on the $\Lambda$ candidate; and the energy and momentum of the $\Lambda$ candidate in the laboratory frame.
The $\psi_D$ mass is specifically excluded from the BDT in order to limit potential bias in the classifier, and the BDT is trained on a signal sample spanning
a wide range of $\psi_D$ masses. The distribution of the BDT score, $\nu_{\rm{BDT}}$, is shown in Fig.~\ref{fig:BDTscore}.
%The optimal criteria on the BDT score are determined by maximizing the Punzi figure-of-merit $\epsilon / (\alpha/2 + \sqrt{N_{\rm bkg}})$~\cite{Punzi01},
%where $\epsilon$ is the signal efficiency, $N_{\rm bkg}$ is the estimated number of background events from the inclusive MC samples, and $\alpha$ is the number of standard
%deviations of desired sensitivity, set to 3.
We select events with a BDT score greater than 0.75, selecting 41 events in the data.
%To further improve the signal purity, we  require a $B_{\rm tag}$ mass between $5.27 <m_{\rm ES} < 5.29 \GeVcc$ and a $\Lambda$ candidate mass in the range $1.110 < m_\Lambda< 1.121 \GeVcc$.
The resulting $\psi_D$ mass distribution is shown in Fig.~\ref{fig:fnlEvts}. Approximately half of the expected
background consists of $e^+e^- \rightarrow q\bar{q}$ events, and the remainder arises from $B\bar{B}$ events.

The signal efficiency varies between $5.9 \times 10^{-4}$ at $m_{\psi_D}=1.0 \GeVcc$ and
$2.1 \times 10^{-4}$ around $m_{\psi_D} = 4.2 \GeVcc$, taking into account the $B_{\rm tag}$ reconstruction efficiency, the selection
criteria, and the $\Lambda \to p\pi^{-}$ branching fraction. As shown in Fig.~\ref{fig:BDTscore}, the inclusive background MC samples do not accurately
reproduce the data. This discrepancy arises from a mis-modeling of several branching fractions used in the simulation, resulting in differences in
$B_{\rm tag}$ reconstruction efficiencies~\cite{hrm05}, as well as differences in charged and neutral particle reconstruction efficiencies, PID efficiencies,
and the modeling of variables used in the BDT. We correct the simulation in a two-step procedure, using sideband data selected with the criteria, described above, applied before the BDT selection, except with the looser requirement $5.20 < m_{\rm ES} < 5.29 \GeVcc$. The region $-0.5<\nu_{\rm{BDT}}<0.75$, largely dominated by
$e^+e^- \rightarrow q\bar{q}$ ($q=u,d,s,c$) events, is used to extract a correction factor for continuum production, $f_{udsc}$, by rescaling the
corresponding MC predictions to the number of observed events. The correction factor for $B\bar{B}$ production, $f_{B^{0}\bar{B}^0}$, is determined from
data in the complementary region $\nu_{\rm{BDT}} < -0.5$, assuming equal contributions from $B^0\bar{B}^0$ and $B^+B^-$. We obtain $f_{udsc}=1.34 \pm 0.10$
and $f_{B^{0}\bar{B}^0} = 1.06 \pm 0.08$. Under the assumption that the $B\bar{B}$ correction factor is independent of the signal $B$ decay mode, we
rescale the signal efficiency by $f_{B^{0}\bar{B}^0}$, and propagate the corresponding uncertainty as a systematic uncertainty.

We extract the signal yield by scanning the $\psi_D$ mass spectrum in steps of the signal mass resolution, $\sigma_m$, probing a total of
193 mass hypotheses. The resolution is estimated by performing fits of a Bukin function~\cite{bukin} to the $\psi_D$ mass distribution for
each signal MC sample, and interpolating the results to the full mass range. The results vary between $90 \MeVcc$ at $m_{\psi_D}=1.0 \GeVcc$ and
$6 \MeVcc$ at $m_{\psi_D} = 4.2 \GeVcc$. The signal yield is determined by counting the number of events in a window of $\pm 3 \sigma_m$ centered
around the $\psi_D$ mass hypothesis. The background is evaluated in two sideband regions of $\pm 3 \sigma_m$ surrounding the signal window, except near $m_{\psi_D} = 4.2~\GeVcc$, where a single region is used.
% Actually, we didn't distinguish the end points, but there is no background entrance already anyway, it is treated as zero background
The largest local significance is found to $2.3 \sigma$, observed near $m_{\psi_D} = 3.7 \GeVcc$, corresponding to a global significance
of $0.4 \sigma$ after including trial factors~\cite{Gross:2010qma}, consistent with the null hypothesis.

In the absence of a signal, upper limits on the branching fraction $B^0 \to \psi_D \Lambda$ are derived at 90\% confidence level (CL)
by applying a profile likelihood method~\cite{rolke} for each $\psi_{D}$ mass hypothesis. The number of signal and background events is assumed to
follow Poisson distributions, while the efficiency is modeled with a Gaussian having a variance equal to the total systematic uncertainty. Systematic
uncertainties arising from track and neutral reconstruction efficiencies, $B_{\rm tag}$ reconstruction efficiencies, selection criteria, and modeling
of the BDT variables are included in the $B\bar{B}$ correction factor described above. Other sources of uncertainty include the
$\Lambda \rightarrow p \pi^-$ branching fraction (0.8\%), the integrated luminosity (0.6\%)~\cite{pepii}, and  the limited statistical precision of the signal
MC samples (0.7-4.6\%). The total systematic uncertainty, obtained by summing in quadrature the different contributions, varies between 7.8 and 9.1\%.

\begin{figure}[htbp]
 \includegraphics[width = 0.47\textwidth]{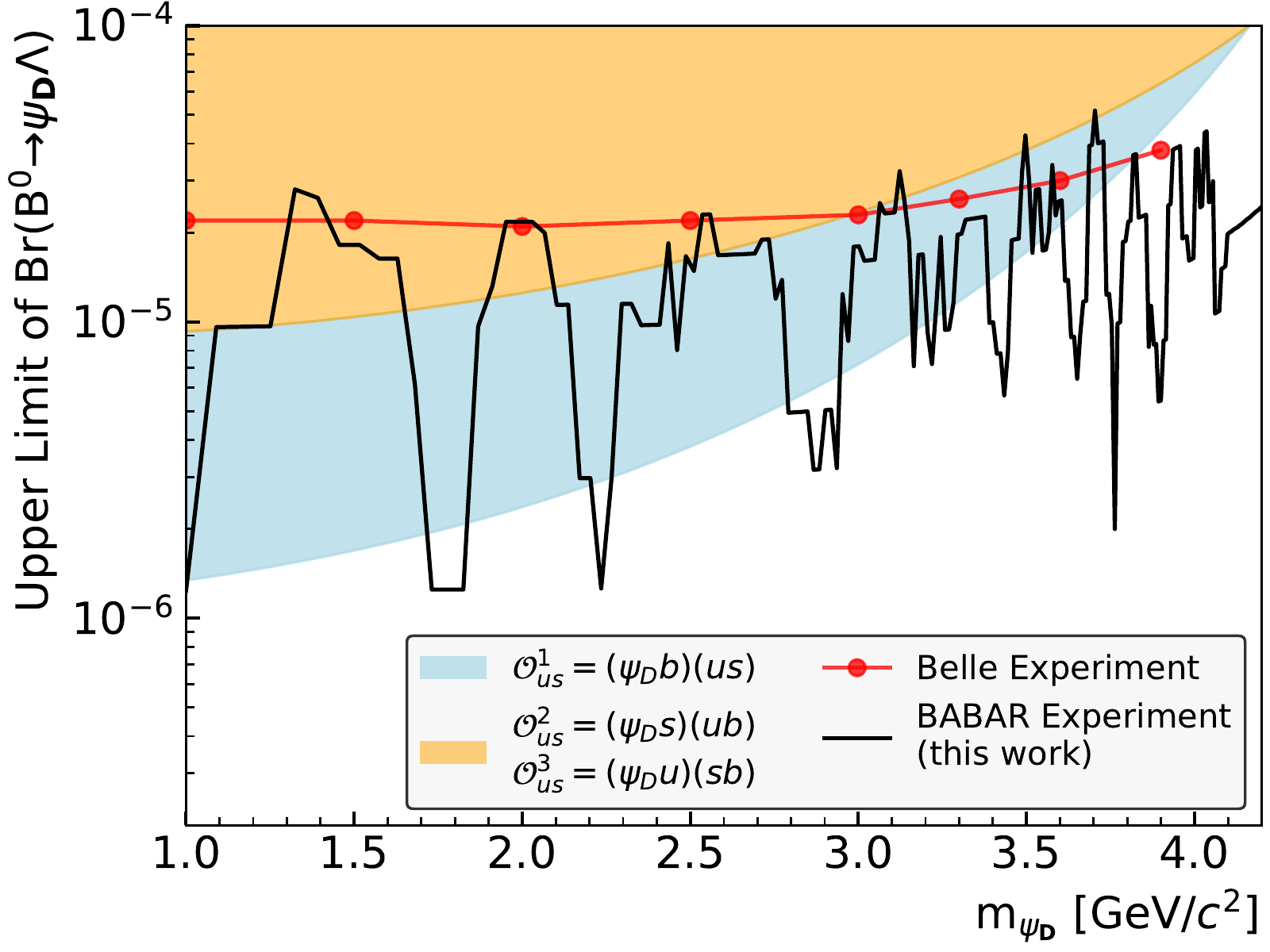}
 \caption{Upper limits on the $B^0 \rightarrow \psi_{D}\Lambda$ branching fraction at 90\% CL, together with previous constraints~\cite{belle}. The light blue and orange (orange only) region shows the values of the $B^0\rightarrow\psi_{D}\Lambda$ branching fraction allowed to successfully generate $B$ Mesogenesis for the $\mathcal{O}^{1}_{us} = (\psi_{D}b)(us)$ ($\mathcal{O}^{2}_{us} = (\psi_{D}s)(ub)$
 and $\mathcal{O}^{3}_{us} = (\psi_{D}u)(sb)$) effective operators~\cite{Alonso-Alvarez:2021qfd}.}
 \label{fig:fnlRes}
\end{figure}

The results are displayed in Fig.~\ref{fig:fnlRes}, together with the previous measurement from the Belle Collaboration and theoretical predictions for different type of
operators generating $B$ Mesogenesis. We probe branching fractions in the range $0.13 - 5.2\times 10^{-5}$, improving previous constraints by up to
an order of magnitude. These bounds exclude most of the remaining parameter space for the $\mathcal{O}^{2}_{us} = (\psi_{D}s)(ub)$ and
$\mathcal{O}^{3}_{us} = (\psi_{D}u)(sb)$ operators, and a significant fraction of the region allowed for $\mathcal{O}^{1}_{us} = (\psi_{D}b)(us)$
operators above $m_{\psi_D} > 2.8 \GeVcc$.

In summary, we report a search for baryogenesis and dark matter in the process $B^0 \to \Lambda \psi_D$ with a fully reconstructed
$B_{\rm tag}$ meson. No significant signal is observed, and upper limits on the branching fraction at the level of $10^{-6} - 10^{-5}$ are
set. These results exclude a large fraction of the parameter space allowed by $B$ Mesogenesis. Future measurements at Belle-II should
be able to fully explore the remaining region.

\acknowledgments
We thank G. Elor and M. Escudero for useful discussions on the B-Mesogenesis mechanism.
We are grateful for the extraordinary contributions of our PEP-II colleagues in achieving the excellent luminosity and machine conditions that have made this work possible. The success of this project also relies critically on the expertise and dedication of the computing organizations that support BABAR. The collaborating institutions wish to thank SLAC for its support and the kind hospitality extended to them.

%=======================================================================

\end{document}